# Inverse design of nano-photonic wavelength demultiplexer with a deep neural network approach

**MENGWEI YUAN, GANG YANG, SHIJIE SONG, LUPING ZHOU, ROBERT MINASIAN, AND XIAOKE. YI**[*]

*School of Electrical and Information Engineering, University of Sydney, Sydney, N.S.W. 2006, Australia*
*xiaoke.yi@sydney.edu.au

**Abstract:** In this paper, we propose a pre-trained-combined neural network (PTCN) as a comprehensive solution to the inverse design of an integrated photonic circuit. By utilizing both the initially pre-trained inverse and forward model with a joint training process, our PTCN model shows remarkable tolerance to the quantity and quality of the training data. As a proof of concept demonstration, the inverse design of a wavelength demultiplexer is used to verify the effectiveness of the PTCN model. The correlation coefficient of the prediction by the presented PTCN model remains greater than 0.974 even when the size of training data is decreased to 17%. The experimental results show a good agreement with predictions, and demonstrate a wavelength demultiplexer with an ultra-compact footprint of 2.6×2.6μm$^2$, a high transmission efficiency with a transmission loss of -2dB, a low reflection of -10dB, and low crosstalk around -7dB simultaneously.

## 1. Introduction

Photonic integrated circuits (PICs), which are compatible with existing complementary metal-oxide-semiconductor fabrication techniques, have found different applications in the fields of optical communications [1], optical computing [2,3], and biomedical apparatus [4]. As the complexity and functionality of the PICs continue to evolve, an effective and efficient inverse design approach that is applicable for different application scenarios is highly desirable [5-7]. Recently, neural networks (NNs) have shown great potential in photonic signal processing in various areas [8-14]. Inverse design using a NN where a trained model is built to predict inverse geometric solutions for desired spectral transmission, has also emerged as a powerful tool for the rapid design of ultra-compact photonic integrated devices with various functionalities [6,7,15-28]. However, due to the non-unique relationship between the geometry and the spectral transmission, the NN-based inverse design for nano-photonic devices is usually hard to converge [27,28]. To solve the non-uniqueness issue, a tandem architecture consisting of two separated networks, an inverse network (design network) and a fixed forward network (spectrum network), has been recently presented [27-32]. Since the forward models in the reported tandem networks are all pre-trained with the input topology structure data sampled from the simulation process, it could potentially form a much larger input domain than those sampled structures used in the pre-training stages. Therefore, networks using a fixed forward model cannot effectively adapt to such data domain shift unless it intensively samples the whole space of the structures to be designed for the pretraining. Even with intensive sampling, the fixed forward model still cannot fully remove the domain shift problem because the outputs from the inverse network are not truly the binary type, but rather a near-binary value between [0,1] due to the requirements of the calculation of the gradient during the training. Consequently, it causes instability issues in the network model and leads to undesired optical performances for the nano-photonic components in the integrated photonic circuits.

In this paper, a pre-trained-combined network (PTCN), where both the forward and inverse models are jointly trained, is proposed to mitigate the domain shift problem. In addition, to reduce the required input parameter numbers, a spectrum generation (SG) model at the inference stage is introduced to generate a target spectrum with a user-defined crossing wavelength point (CWP) for the prediction by the PTCN. The potential of the PTCN model in the inverse design tasks of the PICs is demonstrated by designing a wavelength demultiplexer, which plays a significant role in various applications as a basic building block of PICs [33]. The proposed network achieves a high validation accuracy of around 99%, which is 2% higher than that of the conventional tandem model. Meanwhile, the PTCN shows remarkable stability and robustness to the quantity and the quality of the training data. The correlation coefficient of the prediction by the presented PTCN model remains greater than 0.974, even when the size of training data is decreased to 17%. The wavelength demultiplexer predicted by the PTCN is demonstrated experimentally. The measured results show low crosstalk of -7dB and a low reflection of less than -10dB, agreeing well with the prediction result.

## 1.1 Structure design

To verify the effectiveness of the proposed PTCN model, a 1×2 wavelength demultiplexer is inverse designed on a standard SOI platform with a 220nm top silicon layer, and 2μm buried oxide layer. The three-port device consists of a multimode region with a footprint of 2.6×2.6μm$^2$ and three identical linear tapers to couple light into and out of the 450nm-wide waveguides. Here, the multimode region is chosen as the inverse design region and is discretized into 20×20 square pixels, where each pixel can be switched between two states: silicon square with or without an air hole. The states of the pixels vary the topology pattern of the region represented by a binary matrix (*P*), and this in turn changes the spectra of the device (*T*), which includes the transmission responses at port 1 and 2, and the reflection at the input port. The labeled data (*P, T*) is then used to train the neural network. As a proof of concept, the pixel pitch and the air hole diameter are chosen to be 130nm and 90nm, respectively, both of which are feasible for mass-manufacturing fabrication, including deep UV lithography and etching processes.

## 1.2 Training data collection for arbitrary CWP

Considering a 20×20 binary pattern leads to an ample searching space with $2^{400}$ possible solutions, a modified direct binary search (DBS) algorithm with simulated annealing is adopted to obtain high-quality training data with low transmission loss and high wavelength selectivity. The data collection process for a wavelength demultiplexer at a specific CWP is elaborated as follows. (i) The algorithm randomly generates an initial 20×20 binary matrix; (ii) The spectral responses of the initialized structure are obtained by 3D finite-difference-time-domain (FDTD) simulation (Ansys Lumerical). The simulation results are then used to calculate the original loss utilizing a mean squared error (MSE)-based loss function, which is expressed as

$$L_{MSE} = \frac{1}{3m}\sum_{i=1}^{3m}(N_i^s - N_i^t)^2 \tag{1}$$

where $N_i^s$ and $N_i^t$ are the $i^{th}$ element in simulated and target spectrum sequences respectively. *m* is the number of wavelength points, and the *3m*-length vector represents a total number of wavelength points for the spectral response at port 1, port 2, and reflection; (iii) The algorithm randomly selects and flips a pixel on the current matrix to generate an updated matrix; (iv) To avoid being trapped in local-optimum results, a simulated annealing strategy is adopted. After every γ iteration (γ =200), the flipping change of pixel state is forced to be accepted. Meanwhile, a new iteration starts by returning to Step (iii). Otherwise, the retention of pixel flipping will be determined by the following steps. First, the MSE of the updated structure is evaluated, and then a comparison between the MSEs of the updated and the current binary patterns is conducted. More specifically, the binary pattern with a lower MSE will be taken as the initial value of the next iteration. The process is repeated until the number of iterations is reached, or the MSE meets a pre-defined threshold *ε*, which is set as 0.01.

The aforementioned data collection process is repeated to obtain wavelength demultiplexers operating at different CWPs, from 1500nm to 1600nm, with an increment of 10nm. During the process, the parameters $\gamma$ and $\varepsilon$ are kept as 200 and 0.01, respectively. The whole training set contains 48000 samples. Here, based on the number of iterations in the data collection process, we divide the samples into two groups: 33000 high-performance data utilized for network training and 15000 low-performance data for model robustness evaluation. Specifically, the first 1200 iterations in the data collection process for each CWP target are taken as the low-performance data. Among selected data, the lowest MSE loss is larger than a pre-defined convergence threshold $\varepsilon'$, which is set as 0.075. On the contrary, the high-performance data is selected in the data collection process with an iteration number larger than 1300, where the highest MSE loss in all selected high-performance data is lower than $\varepsilon'$.

*1.3 PTCN architecture*

The inverse NN is pre-trained using the collected high-quality dataset, which takes the existing power spectrum $T$ as the input and predicts the corresponding topology pattern $P'$. Specifically, the network is implemented by nine fully connected layers with 400 nodes in the output layer and 100 nodes in each of the rest layers, respectively. To prevent the vanishing gradient problem, a residual connection is introduced every two layers to create an identical map and reuse the features at the previous layers. As the inverse NN gives a binary output of either "0" or "1", the task can be considered as a multi-class multi-label classification problem. The binary cross-entropy (BCE) loss employed as the loss function of the inverse network is given by

$$L_{BCE} = -\frac{1}{N}[\sum_{i=1}^{N} P_i log(P_i') + (1-P_i)log(1-P_i')] \qquad (2)$$

where $N$ is the number of output dimensions, $P_i$ and $P_i'$ are the ground truth and predicted binary matrix, respectively. Under the guidance of the BCE loss, the inverse network tries to generate the topology pattern, which is as close as possible to the ground truth.

Meanwhile, a forward network is introduced and pre-trained, which takes the existing topology patterns $P$ as the input and approximates the corresponding spectra $T'$. The forward model is based on the fully connected deep neural network structure, consisting of 7 layers with a width of 400 neurons. The mean absolute error (MAE) is used as the loss function of the forward network, which is defined as

$$L_{MAE} = \frac{1}{3m}\sum_{i=1}^{3m} |T_i - T_i'| \qquad (3)$$

where $T_i$ and $T_i'$ are the ground truth and predicted optical transmission spectra array, respectively.

After both models are pre-trained individually, a novel tandem model in an end-to-end manner is constructed to enable the communication between the forward model and the inverse model. As the weights of the pre-trained forward network during the joint-training process are not fixed, the forward model can effectively adapt to the input data generated from the inverse network's outputs during the course, thus obliviating the domain shift problem in conventional tandem architecture. This tandem-like model is fine-tuned using a linear combination of the two-loss functions given by

$$L_{total} = L_{BCE} + \theta L_{MAE} \qquad (4)$$

where $\theta$ is the weight coefficient to balance the prediction ability between the well-performed forward network and the inverse network. To calculate the gradients during the backpropagation process in neural network training, the output of the inverse network is in a format of floating value within the range of [0,1]. Considering that the forward network is trained with samples of the actual binary type (0 and 1), a customized activation function is introduced at the connecting point between the inverse and the forward models. Here, the activation function is set to be a sigmoid-based function given by

$$y = \frac{1}{1+e^{-\beta(x-\alpha)}} \tag{5}$$

where the $\beta$ is the slope coefficient, $\alpha$ is the threshold and $x$ is the output of the inverse network. This function maps the floating number from the inverse model output layer to a value that is extremely close to either 0 or 1 to make the input of the forward network more similar to the training set.

It is noted that the proposed SG model allows the use of a single-specified CWP value, which is an input of the SG model, to generate the *3m*-length target spectrum (*T″*). The generated spectrum *T″* is then utilized in the well-trained PTCN model to produce the required wavelength demultiplexer topology (*P′*), thus reducing the required input parameter numbers significantly. Specifically, the SG model is constructed by six fully connected layers with the Relu activation function to produce a *3m*-length vector. To train the model, the loss function of the SG model is defined as the MAE between the generated power spectrum (*T″*) and the ground truth (*T*).

## 2. Results and analysis

### 2.1 Neural network training results

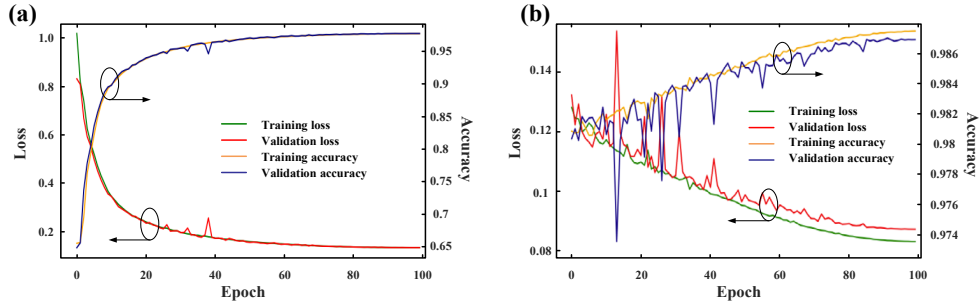

**Fig. 1.** The training loss and the validation accuracy as the functions of epoch number for (a) conventional tandem network and (b) the proposed PTCN model. Training loss: green line; Validation loss: red line; Training accuracy: orange line; Validation accuracy: purple line.

A fast decay learning rate is used and tuned amongst {0.0003, 0.0003, 0.00005, 0.0001} for the pre-trained inverse model, pre-trained forward model, the entire PTCN model, and the SG model, respectively. Both the inverse and the forward networks are pre-trained separately by the same training set with a batch size of 32 before connecting to each other. The 33000 high-performance samples obtained via 3D-FDTD simulation are utilized for training the model with a train-validation ratio of 9:1. Figure 1 shows the convergence process of loss and accuracy for the conventional tandem network and the presented PTCN. As illustrated in Fig. 1(a), the traditional tandem network has a prediction accuracy of around 97%, which means that among 400 topology pixels, there are 12 false predicted pixels. By contrast, our PTCN model exhibits a validation accuracy of around 99%, with only 5 false prediction pixels, as shown in Fig. 1(b).

### 2.2 Neural network evaluation

To test the stability of the PTCN, the size of the training set is varied to 100%, 83%, 67%, 50%, 33%, and 17% of the original dataset volume. Figure 2(a) shows the correlation coefficient between the target and predicted CWP obtained via PCTN and conventional tandem models at different training dataset sizes. In addition, the MAE versus the training data volume for the conventional tandem model and the PTCN model is displayed in Fig. 2(b). It is clear that the PTCN model outperforms the traditional tandem model when the dataset size is reduced. The correlation coefficient of our PTCN model is kept greater than 0.974, showing excellent robustness to the small-size dataset. In contrast, the conventional tandem model struggles to fit the training set, especially when the dataset size is reduced below 50%.

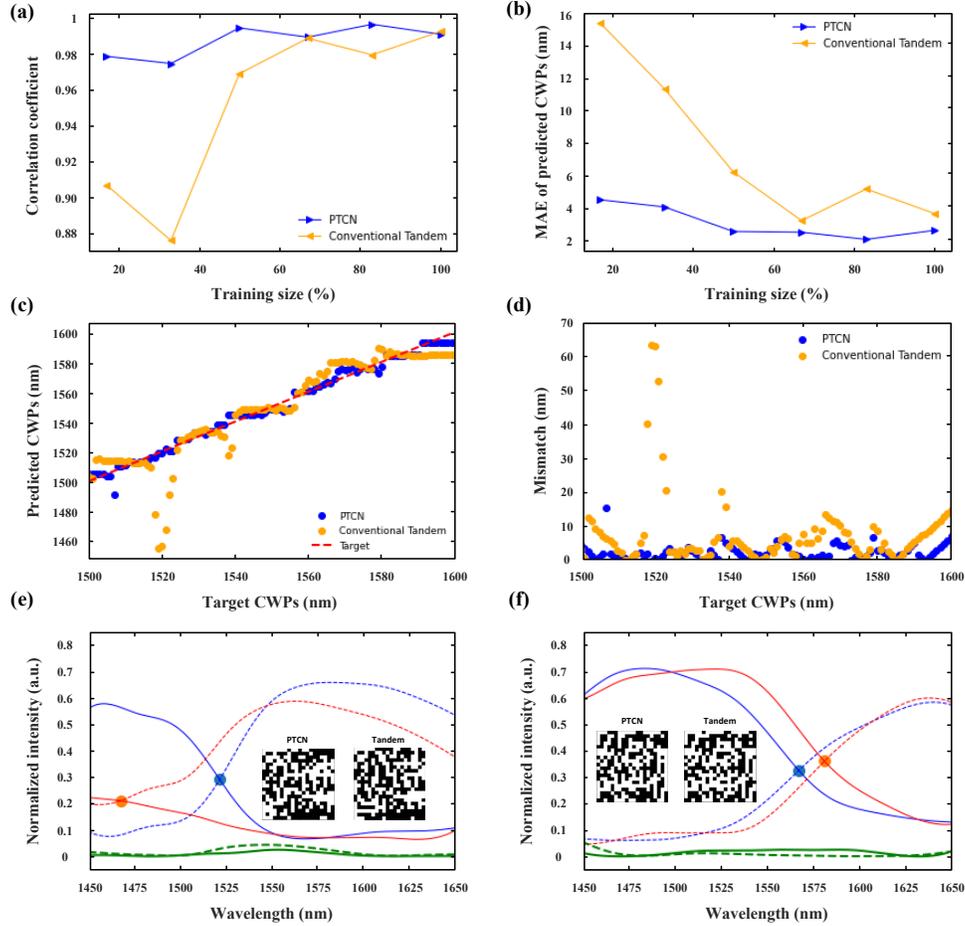

**Fig. 2.** (a) The correlation coefficient between the target and predicted CWP values obtained via PCTN (blue) and convential tandem (yellow) models at different training data sizes. (b) MAEs of the CWP values between the target and predictions obtained via the conventional tandem (yellow) and PTCN (blue) models at different training data sizes. (c) The target and predicted CWP values based on the full dataset of 48000 samples. Target values: red dashed line; PTCN: blue dots; Conventional tandem: yellow dots. (d) CWP mismatch versus the target CWP obtained via PCTN (blue dots) and convential tandem (yellow dots) models. (e-f) Demonstration of the predicted wavelength demultiplexers with the target CWPs of 1521nm (e) and 1566nm (f), where the insets show the predicted topology patterns by PTCN and conventional tandem models. . Optical transmission spectra simulated via the 3D-FDTD at two output ports and the reflection spectrum based on the PTCN and conventional tandem predicted topology patterns. PCTN port 1: blue solid line; Conventional tandem port 1: red solid line; PCTN port 2: blue dashed line; Conventional tandem port 2: red dashed line; PCTN reflection: green solid line; Conventional tandem reflection: green dashed line.

Moreover, to evaluate the robustness of the PTCN model to the quality of training samples, the 15000 noise (low-performance) samples obtained in section 1.2 are added to the dataset. Hence, the total sample number of the new dataset for training both the conventional tandem model and the PTCN model is increased from 33000 to a full dataset of 48000. Figure 2(c) illustrates the comparison between the target and predicted CWPs based on the full dataset, and the CWP mismatch at different target CWPs is plotted in Fig. 2(d). It is clearly seen that the binary pattern predicted by the conventional tandem structure has a larger wavelength mismatch of the CWP values in comparison with the proposed PTCN model. The MAE of the CWP predicted via PTCN model based on the full dataset is reduced from 2.63nm to 2.58nm compared with the result obtained from the original dataset with 33000 training data. In contrast, the MAE of the conventional tandem model is increased from 3.65nm to 8.08nm. Figure2(e-f) demonstrates two wavelength demultiplexers whose target CWPs are 1521nm and 1566nm. The predicted topology patterns are also shown in the inset. It can be seen that the simulated spectral responses of the wavelength demultiplexers predicted by the conventional tandem model have a large difference with the target values showing a degradation in transmission and over 13nm shift of the CWP, while the proposed PTCN model remains superior prediction accuracy and robustness when low-quality data is added.

*2.3 Inverse prediction of wavelength demultiplexer*

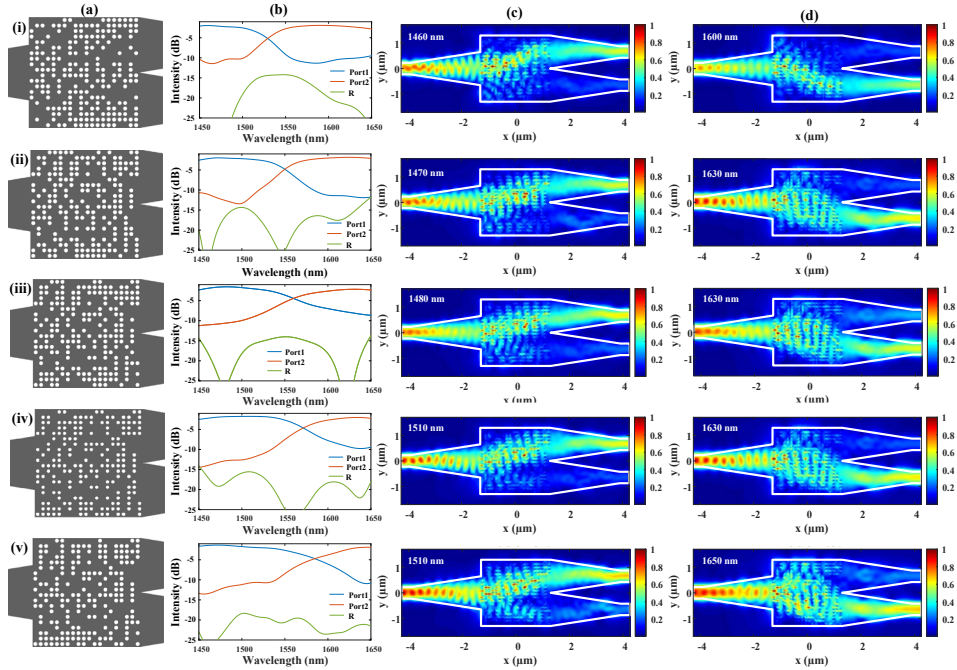

**Fig. 3.** Demonstration of 5 wavelength demultiplexers examples with splitting wavelengths of 1530nm, 1550nm, 1560nm, 1570nm, and 1590nm from (i) to (v). (a) The generated wavelength demultiplexer structure pattern from the PTCN model for different splitting wavelength (Silicon shown in grey and air-hole shown in white). (b) 3D-FDTD simulated transmission spectra at two output ports and the reflection spectrum. (c) The E-field intensity result of predicted structures at the output of port 1 and different CWPs: (i) 1460nm (ii) 1470nm (iii) 1480nm (iv) 1510nm (v) 1510nm. (d) The E-field intensity result of predicted structures at the output of port 2 and different CWPs: (i) 1600nm (ii) 1630nm (iii) 1630nm (iv) 1630nm (v) 1650nm.

To verify the design, wavelength demultiplexer samples with five different CWP targets (1530nm, 1550nm, 1560nm, 1570nm, and 1590nm) are through the well-trained PTCN model. The predicted structure patterns from the PTCN model for different CWPs are shown in Fig. 3(a) (i-v). These are used in the 3D-FDTD simulations to evaluate the optical performance of the predicted samples. The results are presented in Fig.3(b), showing the CWPs of the predicted samples agree well with the target values with less than ±5.3nm offset. This is less than 2.65% of the total simulated wavelength range of 200nm. Moreover, a high transmission efficiency with a transmission loss of around -2dB and low crosstalk of about -8dB can be achieved at passband for both port 1 and port 2 with a reflection power of less than -15dB at the input port for most of the wavelengths. The 3D-FDTD simulated electrical field intensity distribution of TE polarization at port 1 and port 2 of the demultiplexer are presented in Fig. 3(c) and Fig. 3(d), respectively. The E-field intensity at the port 1 and port 2 reveals that the light at different wavelengths is successfully coupled to the target output port, which demonstrates the effectiveness of the predicted wavelength demultiplexer.

*2.4 Fabrication and measurement results*

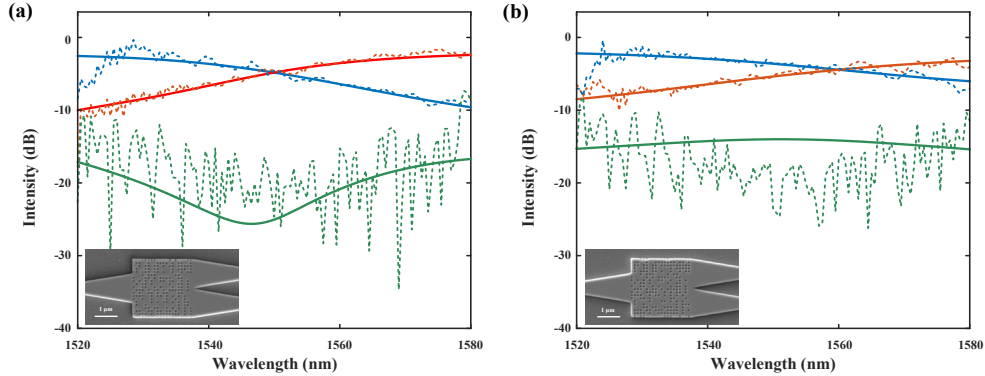

**Fig. 4.** Transmission efficiency of the wavelength multiplexer with target CWPs of (a) 1550nm and (b) 1560nm. The transmission efficiencies at the port 1 and port 2 of the device, as well as the reflection, are represented by blue, red and green colors. The solid lines indicate the simulation results, while the dashed lines display the measured values.

As a proof of concept, two wavelength multiplexers with CWP of 1550nm and 1560nm, were fabricated on a commercially available SOI wafer (Soitec), where the 220-nm-thick silicon waveguide sits on top of a 2μm buried oxide layer and 725μm silicon substrate layer. The device layer with two multiplexers and their respective taper waveguides were patterned via e-beam lithography with optimized proximity effect correction to ensure consistent pixel pitch and etching hole diameter across the entire pattern. The pattern was subsequently etched to a depth of 220nm through an inductively coupled plasma reactive ion etching process. The inset in Fig. 4(a-b) shows the SEM images of the fabricated wavelength multiplexer with a target CWP of 1550nm and 1560nm, respectively. To couple the light between an optical fiber and an SOI waveguide, additional lithography and etching processes were completed to define vertical grating couplers (VGCs) with an etch depth of 70nm. A VGC loop that consists of two paired VGCs was also fabricated for calibration purposes. The measured and predacated transmission efficiency for the fabricated wavelength splitters with CWP of 1550nm and 1560nm are plotted in Fig. 4(a) and Fig. 4(b), respectively, which shows a good agreement with only ±3nm offset between the measured and predicted CWPs. The fabricated demultiplexer exhibits relatively low transmission loss around –2dB and lower than -7dB crosstalk at the passband with a less than -10dB reflection power at the input port for most of the wavelengths.

## 3. Conclusion

In this paper, a novel PCTN for the inverse design of nano-photonic devices has been presented. By creating a joint training process for the inverse and forward models in the neural network, our approach successfully solves the domain shift problem in the conventional tandem architecture. Meanwhile, the proposed model exhibits high stability and robustness to the variation in quantity and quality of training data. The predicted wavelength demultiplexer via

PCTN is experimentally demonstrated, showing a low crosstalk of -7dB and a low reflection of less than -10dB, agreeing well with the prediction result.

**Funding.** The project is supported in part by the Australian Department of Defence.

**Acknowledgments.** The authors would like to acknowledge valuable discussions with Dr. Suen Xin Chew from the University of Sydney.